\documentclass[onecolumn,amsmath,amssymb,superscriptaddress,nofootinbib,11pt,a4paper]{revtex4}
\usepackage{titlesec}
\titleformat{\section}
  {\normalfont\Large\bfseries}  
  {\thesection}                 
  {1em}                         
  {}                            
\titleformat{\subsection}
  {\normalfont\large\bfseries}
  {\thesubsection}
  {1em}
  {}
\pdfoutput=1
\usepackage[T1]{fontenc}
\usepackage{CJK}
\usepackage{xcolor}
\usepackage{amsfonts}
\usepackage{epstopdf}
\usepackage{wrapfig} 
\usepackage{subcaption}
\usepackage{graphicx}  
\usepackage{dcolumn}   
\usepackage{bm}
\usepackage{float}
\usepackage[utf8]{inputenc}
\usepackage{booktabs}
\usepackage{orcidlink}
\usepackage{multirow}

\begin{document}

\title{Repetitive Magnetic Penrose Process}

\author{Ke Wang}
\affiliation{College of Physics and Optoelectronic Engineering, Chongqing Normal University, \\Chongqing 401331, China}
\affiliation{School of Material Science and Engineering, Chongqing Jiaotong University, \\Chongqing 400074, China}

\author{Xin-Yun Hu}
\affiliation{College of Economics and Management, Chongqing Normal University, \\Chongqing 401331, China}

\author{Xiao-Xiong Zeng}
\email{xxzengphysics@163.com}
\affiliation{College of Physics and Optoelectronic Engineering, Chongqing Normal University, \\Chongqing 401331, China}

\begin{abstract}
{The nonlinear iterative repetitive Penrose process  established  recently by Ruffini et al.\  is generalized   to the magnetic Penrose process. The spacetime background is the Kerr metric, the external magnetic field is described by the Wald solution, the electromagnetic field is treated as a test field, and the charge is treated as a test charge. We analyze the optimal conditions for maximum energy extraction on the equatorial plane and find that under these conditions the core mechanism of the iterative process is charging feedback. After charging, the black hole charge is approximately half of the Wald charge. When the neutral parent particle decays, charge conservation forces the infalling fragment to carry charge while the other fragment carries the opposite charge, so the black hole charge accumulates monotonically. The monopole voltage of the Wald solution decreases linearly with the charge $Q$ and crosses zero to reverse sign at a certain point, so the magnetic gain is gradually neutralized by the charge. The repetitive magnetic Penrose process is therefore forced to terminate and eventually falls back to the mechanical Penrose process. The energy extraction efficiency $>100\%$ predicted by the single magnetic Penrose process will therefore be prevented.
}
\end{abstract}
\maketitle

\newpage
\section{Introduction}
Energy extraction from black holes is one of the most fascinating problems in general relativity and black hole thermodynamics. Penrose pointed out in 1969\cite{21} that inside the ergosphere of a rotating black hole, if an incident particle splits into two fragments and the particle on the negative energy orbit falls into the black hole while the other fragment carries positive energy greater than the incident energy and escapes, then the black hole energy is extracted. The energy budget of the black hole is precisely characterized by the Christodoulou--Ruffini irreducible mass\cite{11}, and the extractable energy can reach about $29\%$ of the total mass in the extreme Kerr case. However, the upper limit of single step efficiency  of the mechanical Penrose process is only $20.7\%$\cite{22}, which is low, and it requires the relative velocity of the two fragments to exceed half the speed of light, which is extremely rare in real astrophysical environments\cite{23}. Therefore, it is very difficult for the Penrose process relying purely on mechanical splitting to shoulder the responsibility of explaining high-energy astrophysical phenomena.
In the mid-1980s, studies found\cite{14,32,33,34,35,36} that the negative energy of a charged particle can originate from the interaction between matter and the electromagnetic field, and does not need to originate from a sufficiently large mechanical relative velocity. The magnetic Penrose process was thus born. At present, the magnetic Penrose process has been systematically studied in many spacetime backgrounds\cite{1,2,3,4,5,6,7,8,9,37}. Compared with the mechanical Penrose process, the magnetic Penrose process has many advantages. For example, the efficiency can easily exceed $100\%$, energy extraction does not depend on rapid rotation of the black hole, and it remains efficient at low spin.

On the other hand, a single Penrose process is far from sufficient to exhaust the extractable energy of a black hole, so the idea of repetition naturally arises. Misner, Thorne, and Wheeler proposed as early as 1973\cite{38} that in principle all the extractable energy can be taken out through repeated iterations. Recently Ruffini et al.\ pointed out\cite{13,18} (for related early studies, see\cite{39}) that because the iteration is nonlinear, the mass and angular momentum of the black hole must be updated in each cycle, the irreducible mass will then grow nonlinearly, and any finite number of iterations cannot reduce the extractable energy to zero. The black hole will always retain a locked portion of energy. This nonlinear iterative framework was subsequently generalized to a series of spacetime backgrounds\cite{40,41,42,43,44,45,46,47}. In particular, the electric Penrose process proposed by Denardo and Ruffini\cite{48} can also be iterated. Hu et al.\ found\cite{12} that for a Reissner--Nordstr\"om black hole, a finite number of iterations likewise cannot reduce the charge and extractable energy to zero, similar to the third law of thermodynamics. Related work has been generalized to the Kerr--Newman black hole background\cite{49}.

Existing astronomical observations show that compact objects are almost always immersed in magnetic fields\cite{24,25,26,27,28,29,30,31}. Wald gave in 1974 the exact electromagnetic four potential solution for a Kerr black hole immersed in an asymptotically uniform magnetic field\cite{10}. Because the background is the vacuum Kerr spacetime, this combination exactly satisfies the source free Maxwell equations for any test charge $Q$, and the charge enters only in constant coefficients and not in the metric. This property is a classic result of the Wald solution. When the black hole charge reaches $Q_{\rm Wald}=2aB_0M$, the horizon potential vanishes, which is the so-called Wald charge\cite{10}. To date, all studies of repetitive Penrose processes have targeted the mechanical Penrose process or the electric Penrose process. The magnetic Penrose process, which has the highest efficiency and the broadest astrophysical applications, has never been studied. In particular, in the single magnetic Penrose process the energy extraction efficiency is $>100\%$. Whether this conclusion still holds after considering the iterative process is worth further exploration.

The reason why the repetitive iteration of the mechanical Penrose process and the electric Penrose process can advance rapidly is that their background geometry and electromagnetic field structure are relatively simple. The mechanical Penrose process only needs to update the mass and angular momentum on a similar Kerr background, and the electric Penrose process can treat the charge as a metric parameter in the Reissner--Nordstr\"om or Kerr--Newman background, with the electromagnetic field and geometry being self-consistent. The magnetic Penrose process, however, requires simultaneously handling the Kerr geometry, the magnetic field, the canonical momentum of the charged test particle, and the black hole charge feedback. Although the Wald solution exactly satisfies the source free Maxwell equations for any test charge $Q$, the charge $Q$ is only allowed to appear in the constant coefficients of the Killing combination and cannot enter the metric. Once $Q$ is put into the metric and one switches to Kerr--Newman, the background is no longer Ricci flat and the constant coefficient Killing combination is no longer a Maxwell solution. Therefore, the repetitive iteration must be carried out self-consistently in the nontrivial setting where the geometry is Kerr, the electromagnetic field is the charged Wald solution, and $Q$ is a dynamical variable. Second, single step studies of the magnetic Penrose process mostly adopt local criteria. On the equatorial plane, the radial turning point of a charged fragment in the Wald field necessarily exists, and requiring fragment 2 to escape to infinity would make the iteration fail to close already at the first step. How to define the energy output, the horizon flux, and the work done by the external field under local criteria requires careful distinction between canonical momentum and mechanical momentum. These difficulties have so far prevented the repetitive magnetic Penrose process from being incorporated into the nonlinear iterative framework. To address these problems, this paper takes the Kerr metric as the spacetime background, describes the external magnetic field with the Wald solution, treats the electromagnetic field as a test field, and treats the charge as a test charge. The results show that because the parent particle is neutral, charge conservation forces the infalling fragment to carry charge unidirectionally, so the black hole is gradually charged and $Q$ accumulates monotonically. The monopole voltage $A_t(r)$ in the Wald solution that drives the whole process decreases linearly with $Q$ and crosses zero to reverse sign at $Q\to Q_{\rm crit}=aB_0M=\frac12Q_{\rm Wald}$, and the magnetic gain then turns into magnetic loss. That is, the charge injected by the repetitive process itself eventually neutralizes the monopole voltage that drives it. High efficiency is unsustainable, and the magnetic process will eventually degenerate back to the mechanical Penrose process. This conclusion is consistent with the thermodynamic third law-like behavior of the electric Penrose process\cite{12}. The difference is that the charge is not exhausted but injected, and the final charge of the black hole is approximately half of the Wald charge rather than infinite charging.

The paper is organized as follows. Section II introduces the Kerr black hole under the Wald solution. Section III establishes the equations of motion of charged particles and the conservation laws of a single decay, and gives the iteration scheme of the repetitive magnetic Penrose process. Section IV presents the numerical results, including the single step local efficiency, the charge accumulation during iteration, and the efficiency decay. Section V discusses the related conclusions. Section VI gives the conclusions. In the appendix, the optimal decay condition is proved analytically using the effective potential. This paper adopts the geometric units $G=c=1$ and the metric signature $(-,+,+,+)$.

\section{Kerr Black Hole under the Wald Solution}

This paper considers a Kerr black hole with mass $M$ and spin parameter $a\equiv J/M$, immersed in an external magnetic field that is asymptotically parallel to the black hole spin axis and has strength $B_0$. The electromagnetic field is always treated as a test field and does not enter the spacetime geometry. The geometry is always given by the Kerr metric and is independent of the charge carried by the black hole. In Boyer--Lindquist coordinates $(t,r,\theta,\phi)$, the nonzero components of the Kerr metric are
\begin{equation}
 g_{tt}=-\Big(1-\frac{2Mr}{\Sigma}\Big),\quad
 g_{rr}=\frac{\Sigma}{\Delta},\quad
 g_{\theta\theta}=\Sigma,\quad
 g_{\phi\phi}=\frac{\mathcal{A}}{\Sigma}\sin^2\theta,\quad
 g_{t\phi}=-\frac{2aMr}{\Sigma}\sin^2\theta ,
\end{equation}
where
\begin{equation}
 \Sigma=r^2+a^2\cos^2\theta,\quad
 \Delta=r^2-2Mr+a^2,\quad
 \mathcal{A}=(r^2+a^2)^2-a^2\Delta\sin^2\theta .
\end{equation}
The horizon and the outer boundary of the ergosphere are respectively
\begin{equation}
 r_+=M+\sqrt{M^2-a^2},\quad r_{\rm erg}=M+\sqrt{M^2-a^2\cos^2\theta},
\end{equation}
and on the equatorial plane $\theta=\pi/2$ one has $\Sigma=r^2$ and $r_{\rm erg}=2M$. The angular velocity at the horizon is
\begin{equation}
 \Omega_+=\frac{a}{2Mr_+}.
 \label{eq:OmPhi}
\end{equation}

The electromagnetic four potential when the black hole is immersed in an external magnetic field is given by the Wald solution. For a Kerr black hole carrying a test charge $Q$, it is a constant-coefficient combination of the two Killing vectors $\eta^\mu=\delta^\mu_0$ (timelike) and $\psi^\mu=\delta^\mu_3$ (spacelike and axisymmetric)\cite{10}
\begin{equation}
 A_\mu=\frac{B_0}{2}\,\psi_\mu+\Big(aB_0-\frac{Q}{2M}\Big)\eta_\mu .
 \label{eq:WaldComb}
\end{equation}
Since $\eta_\mu$ and $\psi_\mu$ are Killing vectors and the Kerr spacetime is a vacuum solution ($R_{\mu\nu}=0$), Eq. (\ref{eq:WaldComb}) exactly satisfies the source free Maxwell equations $\nabla_\mu F^{\mu\nu}=0$ for any $Q$, that is, the charge appears only in the constant coefficients and does not enter the metric. In this case, we can treat $Q$ merely as a dynamical variable while keeping the electromagnetic field as an exact solution. Expanding Eq. (\ref{eq:WaldComb}) with the metric components and choosing a gauge such that $A_t(\infty)=0$, we obtain on the equatorial plane the explicit form
\begin{equation}
 A_t(r)=\frac{ab-Q}{r},\quad A_\phi(r)=\frac{B_0}{2}\Big(r^2+a^2-\frac{2Ma^2}{r}\Big)+\frac{aQ}{r}.
\label{eq:Apot}
\end{equation}
Here $b\equiv B_0M$ is the dimensionless magnetic field parameter.

\section{Charged Particle Motion and the Repetitive Magnetic Penrose Process}
\label{sec:formalism}

\subsection{Conserved Quantities and Radial Motion}

Consider a test particle with mass $m$ and charge $q$, and denote $\lambda\equiv q/m$. Its conserved energy and angular momentum are
\begin{equation}
 \hat E\equiv -(u_t+\lambda A_t),\quad
 \hat L\equiv u_\phi+\lambda A_\phi .
 \label{eq:EL}
\end{equation}
Here $\hat E,\hat L=E/m,L/m$. To separate the kinematics from the charge, we introduce
\begin{equation}
 X\equiv -u_t=\hat E+\lambda A_t(r),\quad
 Y\equiv u_\phi=\hat L-\lambda A_\phi(r).
 \label{eq:XY}
\end{equation}
Note that $X,Y$ are functions of $r$, while $\hat E,\hat L$ are conserved along the trajectory. The radial equation for a unit-mass particle on the equatorial plane can be written as
\begin{equation}
 r^4 (u^r)^2=\mathcal{R}(r),\quad\mathcal{R}(r)=\big[X(r^2+a^2)-aY\big]^2-\Delta\big[r^2+(Y-aX)^2\big],
 \label{eq:R}
\end{equation}
and $\mathcal{R}\ge0$ is required. The $t$ and $\phi$ components of the four-velocity are given by
\begin{equation}
 u^t=-g^{tt}X+g^{t\phi}Y,\quad u^\phi=-g^{t\phi}X+g^{\phi\phi}Y,
\end{equation}
and the future-pointing condition is $u^t>0$.

\subsection{Conservation Laws of a Single Decay}

A neutral parent particle (mass $\mu_0$) decays at $r_d$ into fragment 1 (mass $\mu_1$, specific charge $\lambda_1$) and fragment 2 ($\mu_2,\lambda_2$). Charge conservation requires
\begin{equation}
 \mu_1\lambda_1+\mu_2\lambda_2=0\quad\Longrightarrow\quad
 \lambda_2=-\frac{\mu_1}{\mu_2}\lambda_1 .
 \label{eq:charge}
\end{equation}
Since $A_\mu$ takes the same value at the decay point, four-momentum conservation $\mu_0u_0=\mu_1u_1+\mu_2u_2$ reduces in the $X,Y$ variables to a purely mechanical form
\begin{equation}
 \mu_0X_0=\mu_1X_1+\mu_2X_2,\qquad \mu_0Y_0=\mu_1Y_1+\mu_2Y_2 .
 \label{eq:consXY}
\end{equation}
Define
\begin{equation}
 \tilde\mu_i\equiv\frac{\mu_i}{\mu_0}(i=1,2),\quad
 \tilde\mu_1+\tilde\mu_2\le1.
\end{equation}
The equality holds only when $\gamma_{01}=1$, namely no kinetic energy release, and the difference $1-\tilde\mu_1-\tilde\mu_2$ is exactly the mass defect converted into the kinetic energy of the fragments. Then Eq. (\ref{eq:consXY}) reads $X_0=\tilde\mu_1X_1+\tilde\mu_2X_2$. The magnetic field effect does not disappear but returns at the step $\hat E=X-\lambda A_t$ and in the feasibility conditions of the fragments. The rest mass of fragment 2 is determined by the mass-shell condition. From $P_2=P_0-P_1$ one obtains
\begin{equation}
 \mu_2=\sqrt{\mu_0^2+\mu_1^2-2\mu_0\mu_1\gamma_{01}},\quad
 \gamma_{01}\equiv -u_0\!\cdot\! u_1\ge1 .
 \label{eq:mu2}
\end{equation}
This imposes a hard upper bound on the mass of fragment 1
\begin{equation}
 \tilde\mu_1\le \gamma_{01}-\sqrt{\gamma_{01}^2-1}\;\simeq\;\frac{1}{2\gamma_{01}}
 \quad(\gamma_{01}\gg1),
\end{equation}
that is, the larger the relative Lorentz factor between fragment 1 and the parent particle, the smaller the mass that can be brought into a negative energy state.

The kinematic constraints of a single event are that the parent particle can reach $r_d$ from outside and that $\mathcal{R}_0\ge0$ holds for $r\ge r_d$, that fragment 1 is future pointing at $r_d$ and that $\mathcal{R}_1\ge0$ holds all the way to the horizon so that it falls into the black hole, and that fragment 2 is outward and future pointing at $r_d$ with $\mathcal{R}_2>0$.

\subsection{Repetitive Magnetic Penrose Process}

The energy budget of the black hole is described through the Christodoulou--Ruffini irreducible mass\cite{11}. According to the test charge convention of this paper, the geometry is the Kerr metric, so we adopt
\begin{equation}
 M_{\rm irr}=\frac12\sqrt{2Mr_+}=\sqrt{\frac{A}{16\pi}},\quad r_+=M+\sqrt{M^2-a^2}.
 \label{eq:Mirr}
\end{equation}
The extractable energy is
\begin{equation}
 E_{\rm extractable}\equiv M-M_{\rm irr}.
\end{equation}
When fragment 1 is absorbed, the energy and angular momentum fluxes entering the horizon are given on the horizon by the mechanical four-momentum $p_1^a=\mu_1u_1^a$ of the particle (the electromagnetic contribution contained in the canonical momentum $\mu_1u_1^a+\mu_1\lambda_1A^a$ is borne by the external field and is not counted in the black hole parameters) 
\begin{equation}
 \Delta M=-p_{1a}\xi^a\big|_{r_+}=\mu_1X_1(r_+),\quad
 \Delta J=p_{1a}\psi^a\big|_{r_+}=\mu_1Y_1(r_+),\quad
 \Delta Q=\mu_1\lambda_1 ,
 \label{eq:flux}
\end{equation}
where
\begin{equation}
 X_1(r_+)=\hat E_1+\lambda_1A_t(r_+),\quad
 Y_1(r_+)=\hat L_1-\lambda_1A_\phi(r_+) .
 \label{eq:XYH}
\end{equation}
Note that $\Delta M\neq\mu_1\hat E_1$, the latter being the conserved energy at infinity. The difference between the two, $\mu_1\lambda_1A_t(r_+)$, is exactly the work done by the external electromagnetic field on the particle during its fall. In each cycle, the black hole parameters are updated according to
\begin{equation}
 M\leftarrow M+\Delta M,\quad J\leftarrow J+\Delta J,\quad Q\leftarrow Q+\Delta Q,   \label{xxxxxxx}
\end{equation}
and then $a=J/M$, $r_+$, $\Omega_+=a/(2Mr_+)$, $M_{\rm irr}$, $b=B_0M$, and the next-step $A_t,A_\phi$ including the new $Q$ are recomputed. From $A=8\pi Mr_+$ and the Kerr first law $\Delta M-\Omega_+\Delta J=\frac{\kappa_+}{8\pi}\Delta A$, substituting Eq. (\ref{eq:flux}) and using the identity $A_t+\Omega_+A_\phi=aB_0-\frac{Q}{2M}$, one obtains
\begin{equation}
 \frac{\kappa_+}{8\pi}\Delta A
 =\mu_1\big[X_1-\Omega_+Y_1\big]_{r_+}
 =\mu_1\Big[\hat E_1-\Omega_+\hat L_1+\lambda_1\Big(aB_0-\frac{Q}{2M}\Big)\Big]
 =-p_{1a}\chi^a\big|_{r_+}\;\ge\;0 .
 \label{eq:arealaw}
\end{equation}
The rightmost side of Eq. (\ref{eq:arealaw}) is the inner product of a future pointing particle with the future-pointing null generator, so it holds automatically for any fragment 1 that actually reaches the horizon. The constraint that really needs to be imposed during iteration is the nonlocal accessibility of fragment 1, $\mathcal R_1(r)>0\,(r_+<r\le r_d)$, while (\ref{eq:arealaw}) serves as a consistency check. From $\Delta A\ge0$ one obtains $\Delta M_{\rm irr}\ge0$, and the Kerr relation $M^2=M_{\rm irr}^2+J^2/(4M_{\rm irr}^2)$ guarantees that $a\le M$ is maintained automatically.

Denote the energy transported to far infinity in a single event by
\begin{equation}
 E_{{\rm out}}\equiv\mu_2\hat E_2-\mu_0\hat E_0=-\mu_1\hat E_1,
 \label{eq:ffffff}
\end{equation}
which differs from $-\Delta M=-\mu_1X_1(r_+)$ by the external field contribution $\mu_1\lambda_1A_t(r_+)$. The energy return on investment\cite{12}
\begin{equation}
 {\rm EROI}_N\equiv\frac{\sum_{i=1}^{N}E_{{\rm out},i}}{N\mu_0}
 =\frac{1}{N}\sum_{i=1}^{N}\eta_i\,X_{0,i},\quad
 \eta_i\equiv-\frac{\mu_1\hat E_{1,i}}{\mu_0X_{0,i}} ,
 \label{eq:EROI}
\end{equation}
that is, the average of the product of each single-step $\eta_i$ and the parent particle specific energy $X_{0,i}$. The energy utilization efficiency\cite{12}
\begin{equation}
 {\rm EUE}_N\equiv\frac{E_{\rm extracted,N}}
 {E_{\rm extracted,N}+\Delta M_{\rm irr}}
 =\frac{\sum_i(-\Delta M_i)}{\sum_i(-\Delta M_i)+\Delta M_{\rm irr}},\qquad
 \Delta M_{\rm irr}=M_{{\rm irr},N}-M_{{\rm irr},0},
 \label{eq:EUE}
\end{equation}
measures the allocation of irreversibility, that is, of the lost extractable energy, how much is truly output and how much is locked into the irreducible mass. It should be emphasized that $-\Delta M$ rather than $E_{\rm out}$ enters (\ref{eq:EUE}), because the object of Christodoulou--Ruffini accounting is the black hole parameter $M$.

\section{Numerical Calculations}
\label{sec:numerics}

Take $M_0=1$, $b=B_0M_0=0.1$, extreme Kerr $a_*=1$, initial $Q_0=0$, and parent particle specific energy $\hat E_0=1$. This is because in almost all realistic scenarios the incident particle initially moves at nonrelativistic speed\cite{14}. According to the optimal decay condition proved in \ref{bbbbbbb}, all three particles decay at their respective radial turning points. The related analysis will be based on the following facts. (1) Parent particle. The closed-form expression (\ref{eq:Lturn}) directly gives $\hat L_0=\hat L_0^{\rm turn}$, so $V_0=(X_0,Y_0)=(1,\hat L_0^{\rm turn})$, satisfying $\mathcal{B}(V_0,V_0)=1$. (2) Conserved quantities of fragment 1. From $\mathcal{R}_1(r_d)=0$ and the future pointing condition on the upper branch, Eq. \ref{eq:branch}, $X_1$ is uniquely determined by $Y_1$ as
\begin{equation}
  X_1=X_+\big(r_d,Y_1\big) ,
  \label{eq:X1ofY1}
\end{equation}
that is, the upper branch of Eq. (\ref{eq:Xpm}). Therefore fragment 1 has only one free parameter, taken in this paper as $Y_1$, equivalent to the deviation component $s$. (3) Physical conserved quantities. From $X=\hat E+\lambda A_t$ and $Y=\hat L-\lambda A_\phi$ one solves inversely
 \begin{equation}
  \hat E_1=X_1-\lambda_1A_t(r_d),\quad
  \hat L_1=Y_1+\lambda_1A_\phi(r_d) ,
 \end{equation}
and imposes $\hat E_1<0$. (4) Mass fraction. Since $\mathcal{R}_0=\mathcal{R}_1=0$, one has $\gamma_{01}=\mathcal{B}(V_0,V_1)$, so
 \begin{equation}
  \tilde\mu_1^{\max}=\gamma_{01}-\sqrt{\gamma_{01}^2-1} .
 \end{equation}
Since $\eta=\tilde\mu_1|\hat E_1|/\hat E_0$ increases monotonically with $\tilde\mu_1$, we take $\tilde\mu_1=\tilde\mu_1^{\max}$. Equivalently, if one writes $V_1=tV_0+sW$, where $W$ is the $\mathcal{B}$-orthonormal spacelike companion vector, then $t=\sqrt{1+s^2}$ and $\tilde\mu_1^{\max}=\sqrt{1+s^2}-|s|$.

Thus the single step efficiency reduces to a function of a single variable
\begin{equation}
 \eta(s)=\frac{\big(\sqrt{1+s^2}-|s|\big)
 \Big[\lambda_1A_t(r_d)-X_1(s)\Big]}{X_0},\quad
 X_1(s)=\sqrt{1+s^2}\,X_0+s\,X_W ,
 \label{eq:etas}
\end{equation}
where $X_W$ is the $X$ component of the companion vector $W$, completely determined by the metric at $r_d$ and $V_0$. After giving $r_d$ and $\lambda$, this paper performs a one-dimensional scan over $s$ and takes the maximum.

From the above, $\tilde\mu_1=\tilde\mu_1^{\max}$ is equivalent to $\tilde\mu_2=0$, so Table \ref{tab:eta} gives the supremum under each parameter set, approached by the limit $\tilde\mu_2\to0^+$ but not truly attainable. Numerically we take $\tilde\mu_2=10^{-3}$ to compute the outer turning point $r_{\max}$ of fragment 2.

\subsection{Single Step Local Efficiency: Magnetic Gain and $\eta>100\%$}

Table \ref{tab:eta} gives the single step efficiency and its construction parameters for different decay radii and specific charges at $a_*=1$ and $Q=0$. Fragment 2 is only required to move outward at $r_d$ and is not required to escape to infinity\cite{1,2,3,4,5,6,7,8,9,37}. Fragment 1 must truly fall into the horizon, that is, $\mathcal{R}_1(r)\ge0$ and future pointing must hold for all $r\in(r_+,r_d]$. This is a constraint independent of Eq. (\ref{eq:arealaw}), which only guarantees $\mathcal R_1(r_+)\ge0$ at the horizon endpoint.

\begin{table}[h]
\centering
\caption{Single step local efficiency $\eta$ for the three-turning-point configuration ($a_*=1$, $Q=0$, $b=0.1$, $\hat E_0=X_0=1$). The parentheses in each cell give the outer turning point $r_{\max}/M$ of fragment 2 (taking $\tilde\mu_2=10^{-3}$). When $\lambda_1=0$, fragment 2 is uncharged, so $r_{\max}\to\infty$.}
\label{tab:eta}
\begin{tabular}{c|ccccc}
\hline\hline
 $r_d/M$ & $\lambda_1=0$ & $10$ & $20$ & $50$ & $100$ \\
\hline
 1.20 & 0.146 ($\infty$) & 0.297 (5.93) & 0.701 (2.09)
       & 1.424 (1.67) & 1.773 (1.56) \\
 1.35 & 0.109 ($\infty$) & 0.232 (6.03) & 0.535 (2.31)
       & 0.904 (1.95) & 1.016 (1.88) \\
 1.50 & 0.077 ($\infty$) & 0.181 (6.13) & 0.412 (2.53)
       & 0.626 (2.23) & 0.677 (2.17) \\
\hline\hline
\end{tabular}
\end{table}

From Table 1, the efficiency grows significantly with the specific charge and can exceed $100\%$. High efficiency corresponds to strongly localized fragment 2. The $r_{\max}\approx1.6$--$2.0M$ corresponding to $\eta>1$ means that fragment 2 is confined by the magnetic field not far outside the decay point, whereas in the pure mechanical case with $\lambda_1=0$, fragment 2 is uncharged and $r_{\max}\to\infty$. The decay point should be close to the horizon. $A_\phi$ vanishes at $r_+$ and grows rapidly outward, while $A_t$ takes its maximum at $r_+$. The efficiency at $r_d=1.5M$ is systematically lower than at $r_d=1.2M$.

\subsection{Iteration: Charge Accumulation and Neutralization of the Monopole Voltage}

Now the above single step configuration is used for iteration. Take the same parameters as the highest-efficiency cell in Table \ref{tab:eta}, namely $\lambda_1=100$ and $r_d=1.2M$, with the initial state being the standard extreme Kerr parameters, i.e., $a_*^{(0)}=1$, $M_0=1$, $Q_0=0$, $\mu_0=10^{-3}M_0$, and $\hat E_0=1$. In each cycle $\eta$ is maximized, and $(M,J,Q)$ is updated according to Eq. (\ref{xxxxxxx}), after which $r_+$, $\Omega_+$, $M_{\rm irr}$, and the next step $A_t,A_\phi$ are recomputed. The initial state is extreme Kerr, so
\begin{equation}
 r_{+,0}=M_0,\quad
 M_{{\rm irr},0}=\frac{M_0}{\sqrt2}=0.70710678,\quad
 E_{\rm extractable,0}=\Big(1-\frac{1}{\sqrt2}\Big)M_0=0.29289322 .
\end{equation}
Table \ref{tab:iter} gives the cycle-by-cycle tracking for $n=0$--$12$. $E_{\rm out}$ is the energy delivered to far infinity as defined by Eq. (\ref{eq:ffffff}), and its cumulative value is listed in the last column.

\begin{table}[htbp]
\centering\small
\caption{Tracking of the iterative process ($a_*^{(0)}=1$, $b=0.1$, $\lambda_1=100$, $r_d=1.2M$, $\mu_0=10^{-3}M_0$, $\hat E_0=1$). In the first six steps the \emph{cumulative} $M_n>M_0$ (the black hole gains net mass), so $M_0-M_n<0$ and EUE is meaningless (omitted as ---).}
\label{tab:iter}
\begin{tabular}{c|ccccccccc}
\hline\hline
 $n$ & $\eta$ & $M/M_0$ & $M_{\rm irr}/M_0$ & $E_{\rm extractable}/M_0$
 & EROI & EUE & $a_*$ & $Q/M_0$ & $\sum E_{\rm out}/M_0$ \\
\hline
 0  & ---    & 1.000000 & 0.707107 & 0.292893 & ---    & ---    & 1.0000000 & 0       & 0 \\
 1  & 1.7732 & 1.000236 & 0.707414 & 0.292821 & 1.7732 & ---    & 0.9999999 & 0.02009 & $1.7732\times10^{-3}$ \\
 2  & 0.9907 & 1.000282 & 0.707433 & 0.292848 & 1.3819 & ---    & 0.9999999 & 0.03306 & $2.7639\times10^{-3}$ \\
 3  & 0.7194 & 1.000266 & 0.707447 & 0.292819 & 1.1611 & ---    & 0.9999999 & 0.04357 & $3.4833\times10^{-3}$ \\
 4  & 0.5642 & 1.000215 & 0.707458 & 0.292757 & 1.0119 & ---    & 0.9999998 & 0.05267 & $4.0475\times10^{-3}$ \\
 5  & 0.4596 & 1.000142 & 0.707464 & 0.292677 & 0.9014 & ---    & 0.9999997 & 0.06082 & $4.5070\times10^{-3}$ \\
 6  & 0.3826 & 1.000051 & 0.707470 & 0.292581 & 0.8149 & ---    & 0.9999996 & 0.06828 & $4.8896\times10^{-3}$ \\
 7  & 0.3228 & 0.999947 & 0.707477 & 0.292470 & 0.7446 & 0.1246 & 0.9999993 & 0.07519 & $5.2124\times10^{-3}$ \\
 8  & 0.2745 & 0.999833 & 0.707485 & 0.292348 & 0.6859 & 0.3059 & 0.9999990 & 0.08166 & $5.4870\times10^{-3}$ \\
 9  & 0.2345 & 0.999710 & 0.707494 & 0.292216 & 0.6357 & 0.4277 & 0.9999986 & 0.08778 & $5.7214\times10^{-3}$ \\
10  & 0.2005 & 0.999580 & 0.707505 & 0.292076 & 0.5922 & 0.5131 & 0.9999981 & 0.09359 & $5.9220\times10^{-3}$ \\
11  & 0.1712 & 0.999444 & 0.707517 & 0.291927 & 0.5539 & 0.5752 & 0.9999974 & 0.09913 & $6.0932\times10^{-3}$ \\
12  & 0.1464 & 0.999300 & 0.710448 & 0.288852 & 0.5200 & 0.1732 & 0.9999407 & 0.10189 & $6.2396\times10^{-3}$ \\
\hline\hline
\end{tabular}
\end{table}

In each cycle $\Delta Q=\mu_0\tilde\mu_1\lambda_1>0$ accumulates unidirectionally, and the monopole induced voltage is proportional to
\begin{equation}
 A_t(r_d)=\frac{aB_0M-Q}{r_d} ,
\end{equation}
which decreases linearly as $Q$ grows and crosses zero to reverse sign at
\begin{equation}
 Q\to Q_{\rm crit}=aB_0M .
 \label{eq:Qcrit}
\end{equation}
Note that $Q_{\rm crit}=\frac12Q_{\rm Wald}$, differing by a factor of $2$ from the Wald charge $Q_{\rm Wald}=2aB_0M$ at which the horizon potential vanishes\cite{10}. In Table \ref{tab:iter}, $A_t(r_d)$ decreases monotonically from $0.08333$ to $6.3\times10^{-4}$ at step 11. At step 12, $Q=0.10189>Q_{\rm crit}$ and the sign has reversed, so magnetic gain turns into magnetic loss. This is the charge injected by the iterative process itself neutralizing the monopole voltage that drives it. Correspondingly, $\eta$ decays monotonically from $1.773$ to $0.1464$, the latter being exactly the pure mechanical value $0.146$ in the $\lambda_1=0$ column of Table \ref{tab:eta}. After the magnetic channel closes, what remains is the ordinary Penrose process. It must be emphasized that row 12 marks the termination of the magnetic gain, that is, of the magnetic Penrose process, and not the termination of the entire energy extraction process. If iteration continues thereafter, the process will degenerate into the ordinary Penrose process and continue to consume the residual extractable rotational energy $0.2889M_0$ with efficiency $\eta\approx0.146$.

After 12 steps, $a_*$ changes only from $1$ to $0.99994$, and $M$ decreases net by only $7.0\times10^{-4}M_0$, accounting for $0.07\%$ of $M_0$, while the cumulative $\sum E_{\rm out}=6.24\times10^{-3}M_0$ delivered to far infinity is $8.9$ times that amount. The difference $5.5\times10^{-3}M_0$ is paid by the external electromagnetic field. This cumulative output accounts for only $2.13\%$ of the initial extractable energy $0.2929M_0$, and the black hole still retains $0.2889M_0$ of extractable energy when the iteration terminates. The energy budget on the black hole side is completely closed. The extractable energy decreases from $0.292893M_0$ to $0.288852M_0$, of which $3.34\times10^{-3}M_0$ is locked into $M_{\rm irr}$ and only $7.0\times10^{-4}M_0$ appears as a net decrease of $M$. This is exactly the conclusion to be expanded in Section \ref{sec:caveat}. The output of this mechanism mainly comes from the external electromagnetic field, and the black hole acts here as a catalyst rather than an energy source. Its consumption of spin is negligible, with $a_*$ changing by only $6\times10^{-5}$, but it is not completely independent of rotation, because the monopole voltage itself, $Q_{\rm crit}=aB_0M$, originates from the spin $a\neq0$, and when $a\to0$ this voltage disappears and the magnetic channel does not open at all.

EROI, like the efficiency, decreases as the number of iterations increases. EUE is incompatible with the first six steps, where the cumulative mass $M_n>M_0$ and the black hole gains net mass. The numerator $M_0-M_n<0$ in Eq. (\ref{eq:EUE}) is then negative while the denominator is positive, giving $\mathrm{EUE}<0$. Note that the single-step $\Delta M_n$ has already become negative since step 3, but the cumulative mass does not first fall below the initial value until step 7 ($M_7=0.999947$), at which point EUE enters $[0,1]$ and rises to $0.575$ (step 11). At step 12, because $Q>Q_{\rm crit}$ makes absorption significantly irreversible, $\Delta M_{\rm irr}$ jumps and EUE falls back to $0.173$. Therefore Eq. (\ref{eq:EUE}) must be supplemented with the cumulative applicability condition
\begin{equation}
 M_0-M_N=\sum_{i=1}^{N}(-\Delta M_i)>0 ,
 \label{eq:EUEvalid}
\end{equation}
whose single-step equivalent criterion is
\begin{equation}
 \Delta M<0\quad\Longleftrightarrow\quad
 |\hat E_1|>\lambda_1A_t(r_+) ,
\end{equation}
that is, the absolute value of the conserved energy of the swallowed fragment must exceed the electromagnetic energy provided by the external field at the horizon. Otherwise the black hole is not extracted from but gains mass. When $Q>Q_{\rm crit}$ makes $A_t(r_+)<0$, this condition is automatically satisfied and the external field changes from supplying energy to consuming energy.

\subsection{Universality of the Efficiency Decay}

Is the monotonic decrease of $\eta$ and EROI with the cycle number observed in the previous subsection a property only of that particular parameter set? This section gives a negative answer. It is an intrinsic property of this mechanism and gives a total constraint depending on the energy reserve of the external electromagnetic field.

Tables \ref{tab:scan1} and \ref{tab:scan2} give two parameter sets. The first set fixes $a_*^{(0)}=1$, $r_d=1.2M$, $\mu_0=10^{-3}M_0$, and $b=0.1$, and only varies the specific charge $\lambda_1$. Each configuration is iterated until $A_t(r_d)\to0$ truly terminates, and the last row is the last step $N$ before termination. $\eta_1$ represents the efficiency of the first iteration, and $\eta_N$ represents the efficiency of the $N$-th iteration. The second set adopts the adaptive decay radius
\begin{equation}
 r_d=r_++0.2\,(2M-r_+)
 \label{eq:rdadapt}
\end{equation}
to keep the relative position of $r_d$ and the horizon unchanged, and varies the initial spin. $Q_{\rm crit}=aB_0M=a_*bM_0$ varies with the spin. The two rows $a_*^{(0)}=0.5,0.3$ reach the iteration limit in a single step ($N=1$), so there is no $\eta_N$.
\begin{table}[htbp]
\centering\small
\caption{Varying $\lambda_1$ ($a_*^{(0)}=1$, $r_d=1.2M$, $\mu_0=10^{-3}M_0$, $b=0.1$).}
\label{tab:scan1}
\begin{tabular}{c|ccccc}
\hline\hline
 $\lambda_1$ & $N$ & $\eta_1$ & $\eta_N$ & $\sum E_{\rm out}/M_0$ & $Q_N/M_0$ \\
\hline
 10  & 97  & 0.297 & 0.145 & $1.62\times10^{-2}$ & 0.0968 \\
 20  & 31  & 0.701 & 0.145 & $7.16\times10^{-3}$  & 0.0998 \\
 50  & 13  & 1.424 & 0.145 & $6.02\times10^{-3}$  & 0.1006 \\
100  & 12  & 1.773 & 0.146 & $6.24\times10^{-3}$  & 0.1019 \\
200  & 11  & 1.910 & 0.169 & $6.20\times10^{-3}$  & 0.1002 \\
\hline\hline
\end{tabular}
\end{table}

\begin{table}[htbp]
\centering\small
\caption{Varying the initial spin, with the decay radius adaptive according to Eq. (\ref{eq:rdadapt}) ($\lambda_1=100$, $\mu_0=10^{-3}M_0$, $b=0.1$).}
\label{tab:scan2}
\begin{tabular}{c|ccccccc}
\hline\hline
 $a_*^{(0)}$ & $N$ & $\eta_1$ & $\eta_N$ & $\sum E_{\rm out}/M_0$ & $Q_N/M_0$
 & $Q_{\rm crit}/M_0$ & $Q_N/Q_{\rm crit}$ \\
\hline
 1.00 & 12 & 1.773 & 0.146 & $6.24\times10^{-3}$ & 0.1019 & 0.100 & 1.02 \\
 0.99 & 11 & 1.536 & 0.117 & $5.32\times10^{-3}$ & 0.1000 & 0.099 & 1.01 \\
 0.90 & 8  & 1.159 & 0.093 & $3.41\times10^{-3}$ & 0.0911 & 0.090 & 1.01 \\
 0.70 & 3  & 0.995 & 0.066 & $1.21\times10^{-3}$ & 0.0729 & 0.070 & 1.04 \\
 0.50 & 1  & 1.642 & ---   & $1.64\times10^{-3}$ & 0.1000 & 0.050 & 2.00 \\
 0.30 & 1  & 0.586 & ---   & $5.86\times10^{-4}$ & 0.0761 & 0.030 & 2.54 \\
\hline\hline
\end{tabular}
\end{table}

The two tables give three features. (1) The decay is monotonic. For all configurations with $N\ge2$, point-by-point inspection shows that the decrease between adjacent steps in the $\eta$ sequence is nonnegative. (2) The termination of high-spin configurations occurs at $A_t(r_d)\to0$, that is, $Q\to Q_{\rm crit}=aB_0M=a_*\,b\,M_0$. The first four rows of Table \ref{tab:scan2} give $Q_N/Q_{\rm crit}=1.02,1.01,1.01,1.04$, with the last step slightly exceeding the upper limit, as expected. The two rows $a_*^{(0)}=0.5,0.3$ reach the iteration limit in a single step, and one step injects $\Delta Q\simeq\mu_0\lambda_1\simeq0.1\gg Q_{\rm crit}$, directly crossing the sign-reversal point, hence $N=1$. (3) The final efficiency falls back to near the mechanical value for the same $r_d$ with $\lambda_1=0$ ($0.145$ for $r_d=1.2M$, see Table \ref{tab:eta}). After the magnetic gain closes, what remains is exactly the ordinary Penrose process. The final value $0.169$ for $\lambda_1=200$ is slightly higher because this configuration terminates after only 11 steps and $A_t(r_d)$ at the last step has not yet fully vanished.

As a control, take $a_*^{(0)}=1$ and $\lambda_1=100$, but forcibly reset the black hole charge $Q\to0$ after each cycle, equivalent to the surrounding plasma neutralizing the infalling charge. The basic parameters are exactly the same as in Table \ref{tab:iter}. One obtains Eq. (\ref{eq:decay})
\begin{equation}
\begin{aligned}
\text{$Q$ accumulation (Table \ref{tab:iter})}&\quad \eta=1.773,\;0.991,\;0.719,\;0.564,\;0.460,\;\ldots\\
\text{$Q$ frozen}&\quad \eta=1.773,\;1.775,\;1.774,\;1.775,\;1.774,\;\ldots
\end{aligned}
\label{eq:decay}
\end{equation}
When $Q$ is frozen, $A_t(r_d)$ remains stable around $0.0834$, and $\eta$ stays near $1.774$ within 13 steps without any decay. Thus the efficiency decay is entirely caused by charge accumulation and is unrelated to geometry or kinematics. This directly confirms the monopole voltage neutralization mechanism of the previous subsection.

From $\Delta Q_i=\mu_1\lambda_1$ and $E_{{\rm out},i}=\mu_1|\hat E_{1,i}|$, the ratio of the output of each step to its charge increment is
\begin{equation}
 \frac{E_{{\rm out},i}}{\Delta Q_i}=\frac{|\hat E_{1,i}|}{\lambda_1}
 =\frac{aB_0M-Q_{i-1}}{r_d}-\frac{X_{1,i}}{\lambda_1} ,
\end{equation}
where $Q_{i-1}$ is the black hole charge before absorption in that step. If the charging is sufficiently fine, namely $\Delta Q_i\ll Q_{\rm crit}$, the process terminates at $Q\to Q_{\rm crit}=aB_0M$, and the total output can be written as
\begin{equation}
 \sum_i E_{{\rm out},i}
 =Q_{\rm crit}\,\Big\langle\frac{|\hat E_1|}{\lambda_1}\Big\rangle ,
 \label{eq:budget}
\end{equation}
that is, total output $=$ charge budget $\times$ output rate per unit charge.

Table \ref{tab:scan2} also shows that at $a_*=0.3$ there is still $\eta=0.586$, far exceeding the efficiency of the mechanical Penrose process at the same spin. Note that $\eta_1$ is not monotonic in $a_*$. It rebounds to $1.642$ at $a_*=0.5$, higher than $0.995$ at $a_*=0.7$. The reason is that at low spin the optimal point drifts toward the singular limit $\gamma_{01}\to1$, i.e., $\tilde\mu_1^{\max}\to1$ and $\mu_2\to0$, and $\eta$ is amplified by $|\hat E_1|$ itself. Therefore the high $\eta$ of low-spin configurations partly originates from this degenerate limit and is accompanied by reaching the upper limit in a single step ($N=1$), so its physical realizability is weaker than that of the high-spin multi-row iterative configurations.

In summary, the efficiency decay is not a coincidence of a certain parameter set but an inevitable consequence of the chain neutral parent particle $+$ charge conservation $\Rightarrow$ unidirectional charging of the black hole $\Rightarrow$ neutralization of the monopole voltage. As long as the charge is not neutralized by external plasma, decay and termination into the mechanical Penrose process are unavoidable. However, the total output is not hard-constrained by the charge budget. Coarse step size or charge neutralization can both break through Eq. (\ref{eq:budget}), and the real limitation comes from the energy reserve of the external electromagnetic field.

\section{Discussion}
\label{sec:caveat}

\subsection{Energy Extraction Is a Local Process, Not a Global Process}

This paper does not require fragment 2 to escape to infinity but only requires it to move outward at the decay point $r_d$. This choice is not a makeshift measure but is determined by the structure of the Wald field. From Eq. (\ref{eq:Apot}), the charge term $aQ/r$ in $A_\phi$ decays at large radii, while the magnetic field term grows as $\frac{B_0}{2}r^2$. Denote $\alpha\equiv\lambda B_0/2$ and $\beta\equiv\hat L-\frac12\lambda B_0a^2$. Then at large $r$, $Y=\hat L-\lambda A_\phi\to-\alpha r^2+\beta$ and $X\to\hat E$. Substituting into Eq. (\ref{eq:R}) gives
\begin{equation}
 \mathcal{R}(r)\longrightarrow -\frac{\lambda^2B_0^2}{4}\,r^6+\frac{M\lambda^2B_0^2}{2}\,r^5+\Big[\hat E^2-1+\lambda B_0\hat L-\tfrac12\lambda^2B_0^2a^2\Big]r^4 +O(r^3) ,
\end{equation}
where the coefficient of the $r^6$ term is negative, so the turning point of the charged fragment necessarily exists. When $\hat E^2-1$ dominates and the $O(M/r)$ term (i.e., the $r^5$ term above) and the angular momentum correction are neglected,
\begin{equation}
 r_{\max}\simeq
 \frac{2\sqrt{\hat E^2-1+\lambda B_0\hat L-\tfrac12\lambda^2B_0^2a^2}}
 {|\lambda|B_0}.
 \label{eq:rmax}
\end{equation}
In the flat limit $\hat L\to0$ and $a\to0$, this reduces to
\begin{equation}
 r_{\max}\to2\sqrt{\hat E^2-1}/(|\lambda|B_0)=2\,r_{\rm Larmor}  ,
\end{equation}
where
\begin{equation}
r_{\rm Larmor}\equiv p_\perp/(|q|B)=\sqrt{\hat E^2-1}/(|\lambda|B_0)
\end{equation}
is the relativistic Larmor radius\cite{15,16}. On the equatorial plane, a charged particle moving perpendicular to the magnetic field can only gyrate and cannot flow outward radially. Therefore, in the pure Wald field on the equatorial plane, the condition that fragment 2 escapes to infinity cannot be satisfied for any $b$, and the iteration cannot close already at the first step.

After adopting the local criterion, fragment 2 is reflected by the magnetic mirror between $[r_d,r_{\max}]$ and oscillates back and forth, so the cycle can close without requiring infinity. This is exactly the standard practice in the magnetic Penrose process literature\cite{1,2,3,4,5,6,7,8,9,37}. The efficiencies in these works are all local quantities defined at the splitting point or at some finite radius, and do not require the fragments to reach infinity. The $\eta>100\%$ reported in Table \ref{tab:eta} of this paper should be understood in this sense. The column $r_{\max}\approx1.6$--$2.0M$ in Table \ref{tab:eta} is exactly the quantitative manifestation of this locality. The price is clear. The energy is stored in a charged particle oscillating back and forth in the shell $[r_d,r_{\max}]$ and has not been carried away.

\begin{figure*}[htbp]
\centering
\includegraphics[width=\textwidth]{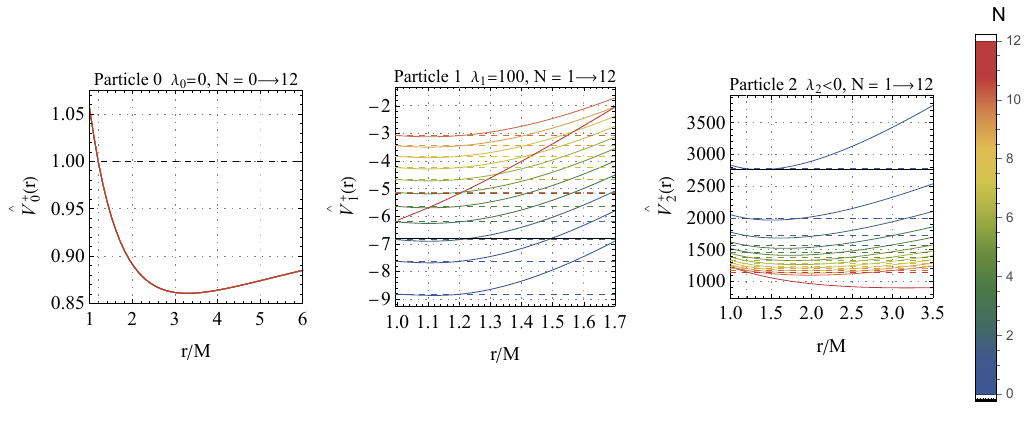}
\caption{Effective potentials $\hat V_i^{+}(r)$ of the three particles as functions of radius, with parameters the same as in Table \ref{tab:iter}.}
\label{fig:1}
\end{figure*}

In Fig. \ref{fig:1} we plot the effective potentials $\hat V_i^{+}(r)$ of the three particles as functions of radius for the case of Table \ref{tab:iter}. The color from blue to red corresponds to the cycle number $N$, and the same color represents the same black hole state. The left panel is the parent particle. The 13 curves correspond to states $N=0\ldots12$, and they all satisfy $\hat V_0^{+}=\hat E_0=1$ at $r_d$, with the 13 curves almost coinciding. The middle and right panels represent fragment 1 and fragment 2. The 12 curves correspond to the $1\ldots12$-th decays, and the dashed line of the same color is the conserved energy $\hat E_i$ of that decay, intersecting the solid line at $r_d$, that is, both are produced at their own turning points. It can be seen that as $Q\to Q_{\rm crit}$, $\hat E_1$ rises from $-8.83$ to $-3.08$ and $\hat E_2$ falls from $2772$ to $1146$, so the magnetic gain is gradually neutralized. At the 12th decay $\hat L_1$ changes sign and the optimal point jumps from the corotating branch to the counter-rotating branch, so the effective potential curve of particle 1 deviates somewhat from those of the first 11 decays. After the 12th decay $Q>Q_{\rm crit}$, the monopole voltage reverses sign and the magnetic channel closes, so there is no 13th curve in the fragment panels. The effective potential of particle 2 clearly reflects the local process of this paper.

\subsection{Source of the Output Energy}

Table \ref{tab:iter} gives a result worth emphasizing. After 12 steps the dimensionless spin of the black hole changes only from $1$ to $0.99994$, and the mass decreases net by $7.0\times10^{-4}M_0$, about $0.07\%$, while the cumulative $\sum E_{\rm out}=6.24\times10^{-3}M_0$ delivered to infinity is $8.9$ times that amount. The difference $5.5\times10^{-3}M_0$ is paid by the external electromagnetic field.

The energy budget of a single step can be written more clearly as a three-party balance. The external field does work $\mu_1\lambda_1A_t(r_+)$ on fragment 1 at the horizon. Part of this work falls into the black hole with fragment 1 , and the rest is output with fragment 2
\begin{equation}
 E_{\rm out}=\underbrace{\mu_1\lambda_1A_t(r_+)}_{\text{paid by external field}}
 \;-\;\underbrace{\Delta M}_{\text{net black hole absorption}} .
\end{equation}
Taking step 1 as an example, the external field pays $\mu_1\lambda_1A_t(r_+)=0.2009\times10^{-3}\times100\times0.1=2.01\times10^{-3}M_0$, the black hole absorbs $\Delta M=\mu_1X_1(r_+)=0.2009\times10^{-3}\times1.172=+2.36\times10^{-4}M_0$, and the net output is $1.77\times10^{-3}M_0$, with the three closing. Note that here $\Delta M>0$. The black hole is not mined but fed. Although fragment 1 has negative conserved energy $\hat E_1<0$, the external field does positive work on it on the way from $r_d$ to $r_+$, so that its mechanical energy upon reaching the horizon, $X_1(r_+)=1.172$, becomes positive.

\section{Conclusions}

This paper generalizes the framework of the nonlinear iterative repetitive Penrose process recently established by Ruffini et al.\cite{13,18} to the magnetic Penrose process. The spacetime background is the Kerr metric, the external magnetic field is described by the Wald solution, and the electromagnetic field is treated throughout as a test field with the black hole charge treated as a test charge. The study shows that the optimal energy extraction condition is that the three particles are simultaneously at their respective radial turning points, which maximizes the efficiency. The single step local efficiency grows significantly with the specific charge and can exceed $100\%$. The high efficiency of the repetitive magnetic Penrose process is unsustainable. The repetitive Penrose process is equivalent to charging the black hole. After charging, the black hole charge is approximately half of the Wald charge, and the magnetic process eventually falls back to the mechanical Penrose process. The total energy output can be written as the product of the charge budget and the output rate per unit charge. The black hole acts almost as a catalyst rather than an energy source in the iterative magnetic Penrose process, because the black hole mass decreases very little and the energy output to far infinity is greater than the decrease of the black hole mass, with the excess coming from electromagnetic energy. In the repetitive magnetic Penrose process, the black hole may even gain mass instead of being extracted from. Although this mechanism consumes almost no spin and the magnetic Penrose process can also proceed at low spin, it is not completely independent of rotation, because the monopole voltage is proportional to the spin, and when the spin tends to zero this voltage disappears and the magnetic channel does not open.

The significance of this study lies not only in providing a numerical example of the repetitive magnetic Penrose process but also in changing the conclusion that the single magnetic Penrose efficiency can easily exceed $100\%$. The high efficiency in single step studies is obtained with fixed black hole parameters and local criteria. Once the black hole charge is allowed to accumulate with the cycles, the monopole voltage decreases linearly and crosses zero to reverse sign at $Q\to Q_{\rm crit}=aB_0M=\frac12Q_{\rm Wald}$, and magnetic gain turns into magnetic loss. Therefore, the repetitive magnetic Penrose process has an intrinsic termination and eventually falls back to the mechanical Penrose process. This termination is not a coincidence of a certain parameter set but an inevitable result of the chain neutral parent particle $+$ charge conservation $\Rightarrow$ unidirectional charging of the black hole $\Rightarrow$ neutralization of the monopole voltage. In addition, the conclusion of this paper is built on local criteria, and the output energy is not truly carried away. If fragment 2 is required to escape to infinity, the iteration cannot close already at the first step. How to make the charged particle oscillating back and forth in the shell truly transport the energy out is a problem worthy of further study. Generalizing the repetitive magnetic Penrose process to exact solutions of magnetized black holes\cite{19,20}, or changing the metric for repetitive magnetic Penrose process studies, is also a future research direction.

\appendix
\renewcommand{\thesection}{Appendix \Alph{section}}
\renewcommand{\theequation}{\Alph{section}\arabic{equation}}
\section{Optimal Decay Condition and Effective Potential Argument}\label{bbbbbbb}
The $\mathcal{R}$ in Eq. (\ref{eq:R}) plays the role of an effective potential. At $\mathcal{R}=0$, $u^r=0$, the radial kinetic energy of the particle vanishes, and this radius is the turning point. $\mathcal{R}\ge0$ is the kinematically allowed region. This section gives an analytical argument for the optimal decay condition using the effective potential.

Rearranging Eq. (\ref{eq:R}) in powers of $X$ and using $\mathcal{A}\equiv(r^2+a^2)^2-a^2\Delta=r^4+a^2r^2+2a^2Mr$, it can be written as the quadratic form
\begin{equation}
 \mathcal{R}=\mathcal{A}X^2-4aMrY\,X-(r^2-2Mr)Y^2-\Delta r^2.
 \label{eq:Rquad}
\end{equation}
Outside the horizon $\mathcal{A}>0$ and $\Delta>0$, so $\mathcal{R}$ is a quadratic function of $X$ opening upward. A quarter of its discriminant can be completely factorized
\begin{equation}
 \frac{  \delta}{4}=(2aMrY)^2+\mathcal{A}\big[(r^2-2Mr)Y^2+\Delta r^2\big]
 =r^2\Delta\big(\mathcal{A}+r^2Y^2\big) .
 \label{eq:disc}
\end{equation}
Since $\mathcal{A}+r^2Y^2>0$, both roots are real and can be written analytically as
\begin{equation}
\mathcal{R}=\mathcal{A}\big(X-X_+\big)\big(X-X_-\big),\quad
 X_\pm(r,Y)=\frac{2aMrY\pm r\sqrt{\Delta}\sqrt{\mathcal{A}+r^2Y^2}}
 {\mathcal{A}},
 \label{eq:Xpm}
\end{equation}
with $X_+>X_-$. The kinematically allowed region is $\mathcal{R}\ge0$, that is, $X\ge X_+$ or $X\le X_-$. Converting Eq. (\ref{eq:Xpm}) back to conserved energy. Since $X=\hat E+\lambda A_t$ and $Y=\hat L-\lambda A_\phi$ both vary with $r$, for given $(\hat E,\hat L,\lambda)$ define the effective potential
\begin{equation}
 \hat V_\pm(r)\equiv X_\pm\big(r,\hat L-\lambda A_\phi(r)\big)-\lambda A_t(r) ,\label{kkkkkkk}
\end{equation}
then
\begin{equation}
 \mathcal{R}=\mathcal{A}\big(\hat E-\hat V_+(r)\big)\big(\hat E-\hat V_-(r)\big) .
 \label{eq:Reff}
\end{equation}
From $u^t=-g^{tt}X+g^{t\phi}Y$ and, on the equatorial plane, $g^{tt}=-\mathcal{A}/(r^2\Delta)$ and $g^{t\phi}=-2aM/(r\Delta)$, one obtains
\begin{equation}
 u^t=\frac{\mathcal{A}X-2aMrY}{r^2\Delta} =\frac{\mathcal{A}}{r^2\Delta}\left(X-\frac{X_++X_-}{2}\right) .
 \label{eq:ut}
\end{equation}
Outside the horizon $\Delta>0$, so the future pointing condition $u^t>0$ is equivalent to $X>(X_++X_-)/2$, that is, $X$ must lie above the midpoint of the two roots. Intersecting this with the two branches of $\mathcal{R}\ge0$, only the upper branch remains
\begin{equation}
 \hat E\;\ge\;\hat V_+(r) .
 \label{eq:branch}
\end{equation}
The lower branch $\hat E\le\hat V_-(r)$ contradicts the future pointing condition and is physically disallowed.

The parent particle is neutral ($\lambda_0=0$), so $X_0=\hat E_0$ and $Y_0=\hat L_0$ are constants, and Eq. (\ref{eq:Reff}) reduces to
\begin{equation}
 \mathcal{R}_0(r)=\mathcal{A}\big(\hat E_0-\hat V_0^+(r)\big)
 \big(\hat E_0-\hat V_0^-(r)\big) .
\end{equation}
As $r\to\infty$, $\hat V_0^\pm\to\pm1$, so incidence from infinity requires $\hat E_0\ge1$, and $\hat V_0^+$ forms a potential barrier. The decay point must satisfy $\mathcal{R}_0(r_d)\ge0$. When equality holds, the parent particle is exactly at its turning point at $r_d$, and then
\begin{equation}
 \hat E_0=\hat V_0^+(r_d)
 \Longleftrightarrow
 \mathcal{A}\hat E_0-2aMr_d\hat L_0
 =+r_d\sqrt{\Delta}\sqrt{\mathcal{A}+r_d^2\hat L_0^2}\;(>0) ,
 \label{eq:turn}
\end{equation}
where the positive sign on the right-hand side locks in the upper branch (future-pointing). Solving for $\hat L_0$ requires squaring Eq. (\ref{eq:turn}), yielding the quadratic equation
\begin{equation}
 \mathcal{D}\hat L_0^2-4aMr_d\mathcal{A}\hat E_0\hat L_0
 +\mathcal{A}\big(\mathcal{A}\hat E_0^2-r_d^2\Delta\big)=0 ,
\end{equation}
whose two roots are
\begin{equation}
 \hat L_0^{\rm turn}=
 \frac{2aMr_d\mathcal{A}\,\hat E_0
 \mp r_d^2\sqrt{\mathcal{A}\Delta}\,
 \sqrt{\mathcal{A}(\hat E_0^2-1)+2Mr_d(r_d^2+a^2)+4a^2M^2}}
 {\mathcal{D}},\\
 \mathcal{D}\equiv4a^2M^2r_d^2-r_d^4\Delta .
 \label{eq:Lturn}
\end{equation}
Using the identity $\mathcal{A}-r^2\Delta=2Mr(r^2+a^2)$, the constant part of the quadratic term can be written as
\begin{equation}
 \mathcal{A}\hat E_0^2-r_d^2\Delta
 =\mathcal{A}(\hat E_0^2-1)+2Mr_d(r_d^2+a^2)>0
 \quad(\hat E_0\ge1) ,
 \label{eq:constpos}
\end{equation}
and the coefficient of the quadratic term can be completely factorized
\begin{equation}
 \mathcal{D}=-\,r_d^2\,(r_d-2M)\big(r_d^3+a^2r_d+2Ma^2\big)
 =r_d^2\,(2M-r_d)\big(r_d^3+a^2r_d+2Ma^2\big) .
 \label{eq:Dfactor}
\end{equation}
The factor in parentheses is always positive for $r_d>r_+$, so the sign of $\mathcal{D}$ is completely determined by $2M-r_d$, and $r=2M$ is exactly the outer boundary of the ergosphere on the equatorial plane, $r_{\rm erg}(\pi/2)$. This leads to completely different properties of the two roots inside and outside the ergosphere.

(i) $r_d>2M$ (outside the ergosphere). $\mathcal{D}<0$. From Eq. (\ref{eq:constpos}), the product of the two roots $\mathcal{A}(\mathcal{A}\hat E_0^2-r_d^2\Delta)/\mathcal{D}<0$, so the two roots have opposite signs, corresponding respectively to $\hat L_0>0$ (corotating) and $\hat L_0<0$ (counter-rotating). Substituting each back into the branch of Eq. (\ref{eq:turn}) with positive right-hand side both gives $\mathcal{A}\hat E_0-2aMr_d\hat L_0>0$, that is, $u^t>0$. Both are future pointing turning points. In this case $(\hat E_0,r_d)$ does not uniquely determine $\hat L_0$.

(ii) $r_+<r_d<2M$ (inside the ergosphere). $\mathcal{D}>0$. In this case, directly treating $\mathcal{R}_0(r_d)=0$ as a quadratic equation in $\hat L_0$
\begin{equation}
 \big(2Mr_d-r_d^2\big)\hat L_0^2-4aMr_d\hat E_0\,\hat L_0
 +\big(\mathcal{A}\hat E_0^2-\Delta r_d^2\big)=0 ,
 \label{eq:Lquad}
\end{equation}
the leading coefficient $2Mr_d-r_d^2>0$ and the constant term is always positive by Eq. (\ref{eq:constpos}), so the product of the two roots is $>0$. Also the sum of the two roots is $=4aMr_d\hat E_0/(2Mr_d-r_d^2)>0$ (for $a>0$), so both roots are positive. In other words, all turning-point solutions inside the ergosphere are corotating, and the counter-rotating turning point $\hat L_0<0$ does not exist as a solution at all. This is exactly the direct consequence of the dragging effect inside the ergosphere making all accessible orbits corotating. Among the two roots of Eq. (\ref{eq:Lturn}), taking the upper sign, that is, the smaller root with $-$ in $\mp$, satisfies Eq. (\ref{eq:turn}) and $u^t>0$. Taking the lower sign, that is, the larger root, gives $\mathcal{A}\hat E_0-2aMr_d\hat L_0=-r_d\sqrt{\Delta}\sqrt{\cdots}<0$, which falls on the lower branch $\hat E_0=\hat V_0^-$ with $u^t<0$ and is past-pointing, so it should be excluded. In this case $\hat L_0$ is unique.

In the degenerate case $r_d=2M$, $\mathcal{D}=0$, the quadratic equation reduces to a linear one, the two branches merge exactly at the ergosphere boundary, and the unique solution is $\hat L_0=(\mathcal{A}\hat E_0^2-\Delta r_d^2)/(4aMr_d\hat E_0)$. All calculations below take $r_d<2M$, belonging to case (ii), so for each given $(\hat E_0,r_d)$, $\hat L_0$ is uniquely determined by Eq. (\ref{eq:Lturn}) with the upper sign, and the parent particle no longer has a free parameter.

From $u^r=\pm\sqrt{\mathcal{R}}/r^2$ and noting the radial orientations of the three particles, namely the parent particle inward ($u_0^r<0$), fragment 1 inward ($u_1^r<0$), and fragment 2 outward ($u_2^r>0$), radial momentum conservation $\mu_0u_0^r=\mu_1u_1^r+\mu_2u_2^r$ becomes (all evaluated at $r_d$)
\begin{equation}
 \mu_0\sqrt{\mathcal{R}_0}=\mu_1\sqrt{\mathcal{R}_1}-\mu_2\sqrt{\mathcal{R}_2} ,
 \label{eq:Rmom}
\end{equation}
from which the strict relation follows immediately
\begin{equation}
 \mu_0\sqrt{\mathcal{R}_0}<\mu_1\sqrt{\mathcal{R}_1} ,
 \label{eq:strict}
\end{equation}
that is, the radial momentum of the parent particle is strictly smaller than that of fragment 1, with the ratio being
$1-\mu_2\sqrt{\mathcal{R}_2}/(\mu_1\sqrt{\mathcal{R}_1})<1$, which tends to 1 when the outward momentum of fragment 2 tends to zero. If $\mathcal{R}_1=0$, then both sides of Eq. (\ref{eq:Rmom}) are simultaneously nonnegative and nonpositive, so necessarily $\mathcal{R}_0=\mathcal{R}_2=0$.

The parent particle must satisfy $\mathcal{R}_0\ge0$ to reach $r_d$, and Eq. (\ref{eq:strict}) further requires its radial momentum to be smaller than that of fragment 1. Taking the inner boundary of this allowed interval, the parent particle decays exactly at its radial turning point, which simultaneously maximizes the margin of Eq. (\ref{eq:strict}) and eliminates the $\hat L_0$ degree of freedom
\begin{equation}
 \mathcal{R}_0(r_d)=0
 \quad\Longleftrightarrow\quad
 \hat E_0=\hat V_0^+(r_d) ,
 \label{eq:best}
\end{equation}
at which point $\hat L_0$ is uniquely determined by Eq. (\ref{eq:Lturn}), and Eq. (\ref{eq:Rmom}) reduces to
\begin{equation}
 \mu_1\sqrt{\mathcal{R}_1}=\mu_2\sqrt{\mathcal{R}_2} ,
\end{equation}
that is, the two fragments carry radial momenta equal in magnitude and opposite in direction. The constraint on fragment 1 is different from that on the parent particle. It is nonlocal, and $\mathcal{R}_1\ge0$ must hold all the way to the horizon, not just $\mathcal{R}_1(r_d)\ge0$.

Next, we will prove that when the energy extraction efficiency is maximal, not only is the radial momentum of the parent particle zero, but the radial momenta of the two daughter particles are also zero.

Define the bilinear form
\begin{equation}
 \mathcal{B}(V,V')\equiv
 -g^{tt}XX'+g^{t\phi}\big(XY'+X'Y\big)-g^{\phi\phi}YY' ,
 \quad V=(X,Y),\ V'=(X',Y') ,
 \label{eq:Bform}
\end{equation}
where $g^{\mu\nu}$ is evaluated at $r_d$. Expanding Eq. (\ref{eq:R}) and comparing with Eq. (\ref{eq:Bform}), one obtains the exact identity
\begin{equation}
 \mathcal{B}(V,V)=1+\frac{\mathcal{R}}{\Delta r^2} .
 \label{eq:BVV}
\end{equation}
The scalar product of the four-velocities of two particles is $\gamma_{01}=-u_0\!\cdot\!u_1=\mathcal{B}(V_0,V_1)-\sqrt{\mathcal{R}_0\mathcal{R}_1}/(\Delta r_d^2)$, where the square-root term comes from the radial part $-g^{rr}u_0^ru_1^r$, whose sign is determined by the radial orientations of the two particles. The parent particle and fragment 1 are both inward, so this term is negative. In particular, when $\mathcal{R}_0(r_d)=0$,
\begin{equation}
 \mathcal{B}(V_0,V_0)=1,\quad \gamma_{01}=\mathcal{B}(V_0,V_1) .
 \label{eq:Bcond}
\end{equation}
The determinant of the matrix of the quadratic form (\ref{eq:Bform}) is $\det\mathcal{B}=g^{tt}g^{\phi\phi}-(g^{t\phi})^2=-1/\Delta<0$ (outside the horizon), so $\mathcal{B}$ is a Lorentzian metric with signature $(1,1)$. This is the pivot of the following proof.

From four-momentum conservation $P_2=P_0-P_1$, one has
$V_2=(V_0-\tilde\mu_1V_1)/\tilde\mu_2$. Using bilinearity and Eq. (\ref{eq:Bcond}),
\begin{align}
 \tilde\mu_2^{\,2}\mathcal{B}(V_2,V_2)
 &=\mathcal{B}(V_0,V_0)+\tilde\mu_1^2\mathcal{B}(V_1,V_1)
 -2\tilde\mu_1\mathcal{B}(V_0,V_1) \nonumber\\
 &=1+\tilde\mu_1^2\Big(1+\frac{\mathcal{R}_1}{\Delta r_d^2}\Big)
 -2\tilde\mu_1\gamma_{01} \nonumber\\
 &=\tilde\mu_2^{\,2}
 +\tilde\mu_1^2\frac{\mathcal{R}_1}{\Delta r_d^2} ,
\end{align}
where the last line has used the mass-shell condition
$\tilde\mu_2^{\,2}=1+\tilde\mu_1^2-2\tilde\mu_1\gamma_{01}$. Dividing both sides by
$\tilde\mu_2^{\,2}$ and using Eq. (\ref{eq:BVV}) for particle 2 gives
\begin{equation}
 \frac{\mathcal{R}_2}{\Delta r_d^2}=\mathcal{B}(V_2,V_2)-1
 =\Big(\frac{\tilde\mu_1}{\tilde\mu_2}\Big)^2
 \frac{\mathcal{R}_1}{\Delta r_d^2} ,
 \quad\text{that is,}\quad
 \mathcal{R}_2=\Big(\frac{\mu_1}{\mu_2}\Big)^2\mathcal{R}_1
 \label{eq:R2R1}
\end{equation}
which shows that the radial momenta of the two fragments are strictly linked. $\mathcal{R}_1=0$ if and only if $\mathcal{R}_2=0$, and it is impossible for one to be at a turning point while the other is not.

Take the $\mathcal{B}$-orthonormal basis $\{V_0,W\}$ satisfying
\begin{equation}
 \mathcal{B}(V_0,W)=0,\quad \mathcal{B}(W,W)=-1 ,
\end{equation}
whose existence is guaranteed by $\det\mathcal{B}<0$. Decompose the conserved quantities of fragment 1 as
\begin{equation}
 V_1=t\,V_0+s\,W ,
 \label{eq:ts}
\end{equation}
where $t=\mathcal{B}(V_0,V_1)=\gamma_{01}$ is the parallel component and $s$ is the deviation component. From Eq.
(\ref{eq:BVV}) and the normalization of $W$,
\begin{equation}
 \mathcal{B}(V_1,V_1)=t^2-s^2=1+\rho ,
 \quad \rho\equiv\frac{\mathcal{R}_1(r_d)}{\Delta r_d^2}\ge0 ,
\end{equation}
so
\begin{equation}
 t=\sqrt{\,1+\rho+s^2\,} .
 \label{eq:tsol}
\end{equation}

Now fix the deviation component $s$ and let $\rho$ increase. From Eq. (\ref{eq:tsol}), $t$ strictly increases with $\rho$, and hence (a) $\gamma_{01}=t$ increases, so $\tilde\mu_1^{\max}=\gamma_{01}-\sqrt{\gamma_{01}^2-1}$ strictly decreases, and (b) from $X_1=tX_0+sX_W$ with $X_0=\hat E_0>0$, $X_1$ strictly increases with $t$, so $|\hat E_1|=\lambda_1A_t(r_d)-X_1$ strictly decreases.

The energy budget of a single cycle is $\mu_0\hat E_0=\mu_1\hat E_1+\mu_2\hat E_2$, so the single-step efficiency
\begin{equation}
 \eta\equiv\frac{\mu_2\hat E_2-\mu_0\hat E_0}{\mu_0\hat E_0}
 =-\frac{\mu_1\hat E_1}{\mu_0\hat E_0}
 =\frac{\tilde\mu_1\big(-X_1+\lambda_1A_t(r_d)\big)}{X_0},
 \label{eq:eta}
\end{equation}
where $\hat E_0=X_0$ (the parent particle is neutral) has been used. The second term in the numerator is the magnetic gain.

The upper bound of the single-step efficiency for given $(\rho,s)$ is
$\eta\le\tilde\mu_1^{\max}|\hat E_1|/\hat E_0$. By (a) and (b), this upper bound, as a function of $\rho$, is the product of two positive functions decreasing in the same direction, and is therefore strictly monotonically decreasing. Hence
\begin{equation}
 \left.\frac{\partial\eta}{\partial\mathcal{R}_1(r_d)}\right|_{s}<0 ,
\end{equation}
and the maximum lies at the endpoint of the allowed interval $\rho=0$, that is,
\begin{equation}
 \mathcal{R}_1(r_d)=0\quad\Longleftrightarrow\quad
 \mathcal{R}_2(r_d)=0 .
 \label{eq:bestfrag}
\end{equation}
In other words, the two fragments should also be produced at their respective turning points. For fragment 1, $r_d$ is its outer turning point ($\mathcal{R}_1>0$ on the inside, so it can fall into the horizon). For fragment 2, $r_d$ is its inner turning point ($\mathcal{R}_2>0$ on the outside, so it can accelerate outward). The two are not contradictory, because $\mathcal{R}$ is a function of $r$, and the same radius can be a turning point of different types for different particles.

It should be emphasized that the supremum is not attainable. In Eq. (\ref{eq:eta}), $\tilde\mu_1\le\tilde\mu_1^{\max}$ takes equality if and only if $\tilde\mu_2=0$, which can be seen from the coincidence of the two roots of $\tilde\mu_2^{\,2}=1+\tilde\mu_1^2-2\tilde\mu_1\gamma_{01}$. Therefore the supremum of $\eta$ is reached only in the joint limit $(\tilde\mu_2,\mathcal{R}_1,\mathcal{R}_2)\to(0,0,0)$. At this point fragment 2 tends to a massless particle, $\lambda_2=-\tilde\mu_1\lambda_1/\tilde\mu_2\to\infty$ while its charge $q_2=-\mu_1\lambda_1$ remains finite, and $r_{\max}\to r_d$. Any configuration with $\tilde\mu_2>0$ can only approach but not reach this upper bound.

In summary, under the optimal energy extraction condition satisfying maximum efficiency, the radial momenta of the three particles are zero. This conclusion is the same as in Refs.\cite{12,13}.

\noindent {\bf Acknowledgments}

\noindent
This work is supported by the National Natural Science Foundation of China (Grants Nos. 12375043, 12575069), Chongqing Normal University Fund Project (Grant No. 26XLB001), and Science and Technology Research Program of Chongqing Municipal Education Commission (Grant No. KJZD-K202600503).

\end{document}